\documentstyle[aps,preprint,floats,epsfig]{revtex}
\begin{document}
\tighten

\title{Multiple Parton Scattering in Nuclei: Twist-Four 
Nuclear Matrix Elements
and Off-Forward Parton Distributions}


\author{Jonathan Osborne and Xin-Nian Wang }
\address{Nuclear Science Division, MS 70-319, Lawrence Berkeley 
National Laboratory\\
Berkeley, CA 94720}

\date{LBNL-50019,April 2002}

\maketitle

\begin{abstract}
Multiple parton scatterings inside a large nucleus generally
involve higher-twist nuclear parton matrix elements. The
gluon bremsstrahlung induced by multiple scattering
depends not only on direct parton matrix elements but
also on momentum-crossed ones, due to the Landau-Pomeranchuk-Migdal
interference effect. 
We show that
both types of twist-four nuclear parton matrix elements
can be factorized approximately into the product 
of twist-two nucleon matrix elements in the limit of 
extremely large nuclei, $A\rightarrow \infty$, as assumed 
in previous studies.  Due to the 
correlative nature of the twist-four matrix elements under
consideration, it is actually the off-forward parton distributions
that appear naturally in this decomposition, rather than the 
ordinary diagonal distributions probed in deeply-inelastic scattering. 
However, we argue that the difference between these
two distribution classes is small in certain kinematic regimes.
In these regions, the twist-four
nuclear parton matrix elements are evaluated numerically and 
compared to the factorized form for different nuclear sizes
within a schematic model of the two-nucleon 
correlation function.
The nuclear size dependence is found to be $A^{4/3}$
in the limit of large $A$, as expected.
We find that the factorization is reasonably good when the momentum
fraction carried by the gluon field is moderate. 
The deviation can be more than a factor of 2, however, for small gluon
momentum fractions, where the gluon distribution is very large.  
\end{abstract}

\pacs{xxxxxxxx}

\narrowtext

\newcommand{\lsim}
{\ \raisebox{2.75pt}{$<$}\hspace{-9.0pt}\raisebox{-2.75pt}{$\sim$}\ }
\newcommand{\gsim}
{\ \raisebox{2.75pt}{$>$}\hspace{-9.0pt}\raisebox{-2.75pt}{$\sim$}\ }

\section{Introduction}

The success of perturbative QCD (pQCD) in describing hard processes in 
high-energy collisions is mainly attributed to the asymptotic freedom
of QCD \cite{Gross,Politzer} at short distances and to
factorization theorems\cite{mueller}. Specifically, the cross sections of
processes that involve large momentum transfers can be factorized
into a convolution of perturbative hard scattering 
cross sections 
and nonperturbative parton distributions and fragmentation
functions that contain long distance physics. Even though they are
not calculable 
within pQCD, these parton distributions and fragmentation functions can 
be rigorously defined in QCD independently of any specific process 
and measured in many different experiments. Such factorization has
been proven up to next-to-leading twist (twist-four) \cite{QS} for hard
processes involving both hadrons and nuclei. We will refer to this as
the generalized factorization.

The leading twist-four contributions in hard processes in nuclei 
normally involve multiple scattering with partons from different 
nucleons. They generally depend on twist-four nuclear 
parton matrix elements
such as
\begin{eqnarray}
\int \frac{dy^{-}}{2\pi}\, dy_1^-dy_2^- 
&&e^{ix_1p^+y^-+ix_2p^+(y_1^--y_2^-)} \theta(-y_2^-)\theta(y^--y_1^-) 
\nonumber \\
&\times& \frac{1}{2}\langle A | \bar{\psi}_q(0)\,
\gamma^+\, F_{\sigma}^{\ +}(y_{2}^{-})\, F^{+\sigma}(y_1^{-})\,\psi_q(y^{-})
| A\rangle\;\; , \label{eq:tw4-1}
\end{eqnarray}
which describes the quark-gluon correlation in a nucleus.
This matrix element also appears in both 
lepton-nucleus deeply-inelastic 
scattering (DIS) \cite{LQS} and in  
Drell-Yan cross section of $pA$
collisions \cite{guo97,fssm}. We will work
in the infinite momentum frame, where the four-momentum 
of the virtual photon and the nucleus (atomic number $A$) 
have the form
\begin{eqnarray}
  q& =&[-Q^2/2q^-, q^-, \vec{0}_\perp], \label{eq:frame} \nonumber \\
  p_A& = & A[p^+,0,\vec{0}_\perp],
\end{eqnarray}
respectively. The Bjorken variable is then $x_B=Q^2/2p^+q^-$.
Our convention for four-vectors is $k^\mu=[k^+,k^-,\vec k_\perp]$, where
\begin{equation}
k^+\equiv {k^0+k^3\over\sqrt{2}}\;\; ,\qquad\qquad
k^-\equiv {k^0-k^3\over\sqrt{2}}\;\; .
\end{equation}

Assuming that 
the two gluon fields in the rescattering process associated 
with Eq.~(\ref{eq:tw4-1}) come 
from the same nucleon in the nucleus due to color confinement,
it has been argued \cite{LQS} that the above twist-four nuclear 
matrix elements
are enhanced by a factor of $A^{1/3}$ as
compared to the leading twist quark distributions in a nucleus,
\begin{equation}
f_q^A(x)=\int \frac{dy^{-}}{2\pi} e^{ixp^+y^-}
\frac{1}{2}\langle A | \bar{\psi}_q(0)\,
\gamma^+\,\psi_q(y^{-})
| A\rangle \, , \label{eq:tw2}
\end{equation}
for $A\gg 1$.
For processes involving a large transverse momentum scale 
$\ell_T^2 \gsim Q^2$, the ratio of the
twist-four contribution and the leading twist one is therefore
proportional to $\alpha_s A^{1/3}/\ell_T^2$.
For large values of $A$,
where the above analysis is valid, this quantity can be related
to an expansion parameter.  
In this sense, the above matrix element is the leading 
higher-twist contribution to hard processes
involving multiple parton scattering in nuclei.

In a recent study \cite{GW}, Guo and Wang extended the generalized
factorization approach to the problem of parton energy loss and modified 
quark fragmentation in DIS off a nuclear target due to gluon bremsstrahlung
induced by secondary quark-gluon scatterings. Because of the 
Landau-Pomeranchuck-Midgal (LPM)\cite{LPM} interference effect,
gluon bremsstrahlung with small transverse momentum, or large
formation time ($\tau_f \sim Q^2/M\ell_T^2$ in the nucleus rest
frame, where $M$ is the nucleon mass), is suppressed.  This limits the 
available phase space of the transverse momentum 
to $\ell_T^2\gsim Q^2/MR_A\sim Q^2/A^{1/3}$, ensuring
the validity of the leading logarithmic approximation in the study
of jet fragmentation for $\ell_T^2\ll Q^2$ in a large
nucleus with $A^{1/3}\gg 1$. The twist-four contribution to
the modified fragmentation function in this case is proportional 
to $\alpha_s A^{2/3}/Q^2$, which depends quadratically on the 
nuclear size. Such a novel quadratic nuclear size dependence has
recently been verified by the HERMES experiment \cite{hermes}.

Similar to other twist-four processes in a nucleus, the nuclear
modification to the fragmentation function is also
proportional to twist-four nuclear 
parton matrix elements. The LPM interference 
effect is explicitly embedded in the combined twist-four parton matrix 
elements.  The quadratic nuclear size dependence of the
modification to the fragmentation function is based on a generalized
assumption that the twist-four parton matrix elements factorize
into twist-two parton distributions in nucleons \cite{LQS}. The
same approximation has been assumed for the momentum-crossed twist-four
parton elements of a nucleus. This is a crude assumption at best, and 
does not specify the condition of validity nor provide any insight 
into the the relationship between nuclear and nucleonic parton 
distributions, which should depend on the nucleon wavefunction 
inside a nucleus.

It is difficult to determine the validity of this approximation
within the framework
of pQCD since nucleons do not appear explicitly in this theory.
However, many hybrid models which employ nonrelativistic 
quantum mechanics to define a wavefunction for nucleons
in a nucleus have been developed in the literature 
\cite{standard,hybrid,newref}.
In these models, the wavefunction allows one to decompose 
nuclear parton distributions into nucleonic ones. 
Phenomenologically, these models assert that scattering processes
involving nuclei can be understood as weighted averages
over scattering processes involving nucleons. Interactions
among nucleons are reflected in the wave function. 
The simplest of these models \cite{convol} arrives at the relation
\begin{equation}
f_{a/A}(x)=A\int_x^A\,\frac{d\alpha}{\alpha}\,\rho(\alpha)
\,f_{a/N}(x/\alpha)
\label{conv}
\end{equation}
between the leading-twist distribution of $a$-type partons in a 
nucleus of size $A$ and the same distribution
found in a nucleon.  The correspondence 
is made via a light-cone nucleon density 
function $\rho(\alpha)$, which is the probability 
of finding a nucleon in a nucleus with longitudinal momentum 
fraction $\alpha$, normalized to 1.  Experimentally, 
Eq.(\ref{conv}) is approximately satisfied for $x\gsim 0.1$. 
At smaller values of $x$, the phenomenon of shadowing 
prevents realization of this naive model \cite{strikman,huang}.

For twist-four matrix elements, one expects 
similar results.  As twist-four
objects are associated with partonic correlations, 
one expects two types of contributions : 
effects associated with
partonic correlations within a single nucleon and 
those associated with 
nucleonic correlations within the nucleus.
The former effects, which involve nucleonic 
twist-four distributions, 
are a simple extension of the convolution model;
one simply substitutes new distributions for the
twist-two ones.  Contributions 
from nucleonic correlatons imply multiple scattering
within the nucleus, and are inherently new effects.
In particular, the distribution $\rho(\alpha)$ in Eq.(\ref{conv})
will be replaced by a more complicated distribution 
describing two-nucleon correlations, or the momentum-sharing between nucleons
in the nucleus.  Since these latter effects involve 
two nucleons, they will be enhanced by factors
of the nuclear radius relative to the nucleonic higher-twist 
effects.  This makes them the dominant contributions
in the limit of large $A$.  

The purpose of this paper is to show that an analysis of
these effects in the spirit of the convolution model
reveals contributions from a more general class of
twist-two matrix elements.  Specifically, the 
decomposition necessarily contains off-forward 
parton distributions (OFPD's) \cite{ji,ji2} rather than 
just the simple diagonal matrix elements.
At present, deeply-virtual Compton scattering (DVCS) is
the only known process which involves
the OFPD's explicitly \cite{ji2}.   We will demonstrate that these
elusive objects could in principle 
also be probed in multiple parton scattering
processes in a nucleus.

This paper is organized as follows.  
In Section II we will give a brief review of multiple parton scattering in
nuclei and modified fragmentation functions, focusing on their relation to
the twist-four parton matrix elements in nuclei. Section III presents 
the main result of this paper, in which we derive a relation between 
a twist-four nuclear matrix element and a convolution 
of two nucleonic OFPD's.  In Section IV, we discuss the 
properties of our result in certain limits.  We will find
that an analytic relationship between the simple factorized 
expression mentioned above and our result is not obvious, but 
numerical models show that they produce the same
results in the limit of sharply-peaked nuclear wave functions.
Section V contains some discussion about and 
conclusions of our results.

\section{Multiple parton scattering}  

DIS on a nuclear target is the simplest environment in which to study
the problem of multiple parton scattering in a nucleus. In
this case, a quark is struck by an energetic virtual
photon and then scatters again with partons from other
nucleons inside the nucleus. The rescattering will induce gluon
bremsstrahlung by the propagating quark and cause the leading
quark to lose energy. Such radiative energy loss will be
manifested in the modification of the quark fragmentation
function as compared to the one measured in DIS off a nucleon target,
where there is no such rescattering. The gluon bremsstrahlung
will interfere destructively with the final-state radiation of
the quark-photon scattering.  This LPM interference effect
will give rise to some
novel nuclear effects in the modified quark fragmentation
function. 

Applying the generalized factorization of twist-four processes to the
exclusive process of hadron production in DIS on a nuclear target
and performing a collinear expansion with respect to initial
parton transverse momentum, one can obtain an effective modified
quark fragmentation function with leading 
higher-twist contributions \cite{GW}:
\begin{eqnarray}
\widetilde{D}_{q\rightarrow h}(z_h,\mu^2)&\equiv& 
D_{q\rightarrow h}(z_h,\mu^2)
+\int_0^{\mu^2} \frac{d\ell_T^2}{\ell_T^2} 
\frac{\alpha_s}{2\pi} \int_{z_h}^1 \frac{dz}{z}
\left[ \Delta\gamma_{q\rightarrow qg}(z,x,x_L,\ell_T^2) 
D_{q\rightarrow h}(z_h/z) \right. \nonumber \\
&+& \left. \Delta\gamma_{q\rightarrow gq}(z,x,x_L,\ell_T^2)
D_{g\rightarrow h}(z_h/z)\right] \, . \label{eq:dmod}
\end{eqnarray}
Here, $D_{q\rightarrow h}(z_h,\mu^2)$ is the usual renormalized
twist-two 
quark fragmentation function in vacuum that satifies the normal
DGLAP \cite{dglap} QCD evolution equation. The additional terms
are the leading higher-twist contributions from multiple parton
scattering and induced gluon bremsstrahlung. These contributions
are very similar in form to the normal gluon radiation in vacuum
except that the modified splitting functions,
\begin{eqnarray}
\Delta\gamma_{q\rightarrow qg}(z,x,x_L,\ell_T^2)&=&
\left[\frac{1+z^2}{(1-z)_+}T^A_{qg}(x,x_L) + 
\delta(1-z)\Delta T^A_{qg}(x,\ell_T^2) \right]
\frac{C_A2\pi\alpha_s}
{\ell_T^2 N_c f_q^A(x)}
\label{eq:dsplit1}\;\; ,\\
\Delta\gamma_{q\rightarrow gq}(z,x,x_L,\ell_T^2) 
&=& \Delta\gamma_{q\rightarrow qg}(1-z,x,x_L,\ell_T^2) 
\label{eq:dsplit2}\;\; ,
\end{eqnarray}
depend on the twist-four two-parton correlation function
\begin{eqnarray}
T_{qg}^A(x,x_T,x_L)&\equiv&
  \int \frac{dy^- }{4\pi} dy^-_1 dy^-_2 \theta(-y^-_2)\theta(y^--y^-_1)
 (1-e^{ix_Lp^+(y^-_1-y^-)})(1-e^{-ix_Lp^+y^-_2}) \nonumber \\
& & \qquad\times e^{i(x+x_L)p^+y^-+ix_Tp^+(y_1^--y_2^-)}
\langle A|\bar{\psi}_q(0)\gamma^+ 
F_\sigma^{a\,\,+}(y^-_2)
F^{+\sigma}_a(y^-_1)
\psi_q(y^-)
|A\rangle, \label{eq:t4}
\end{eqnarray}
where $x_L=\ell_T^2/2p^+q^-z(1-z)$ and 
$x_T\equiv \langle k_T^2\rangle/2p^+q^-z$. 
The virtual corrections supply the 
$\delta$-function contribution to the `+'-function, along with the 
explicit end point contribution
\begin{eqnarray}
\Delta T^A_{qg}(x,\ell_T^2) &\equiv& 
-\int_0^1 dz\,{1+z^2\over(1-z)_+}\, T^A_{qg}(x,x_L)\nonumber\\
&=&\int_0^1 dz\frac{1}{1-z}\left[ 2 T^A_{qg}(x,x_L)|_{z=1}
-(1+z^2) T^A_{qg}(x,x_L)\right] \, , \label{eq:virtual}
\end{eqnarray}
required for conservation of quark flavor.

\begin{figure}
\centerline{\psfig{file=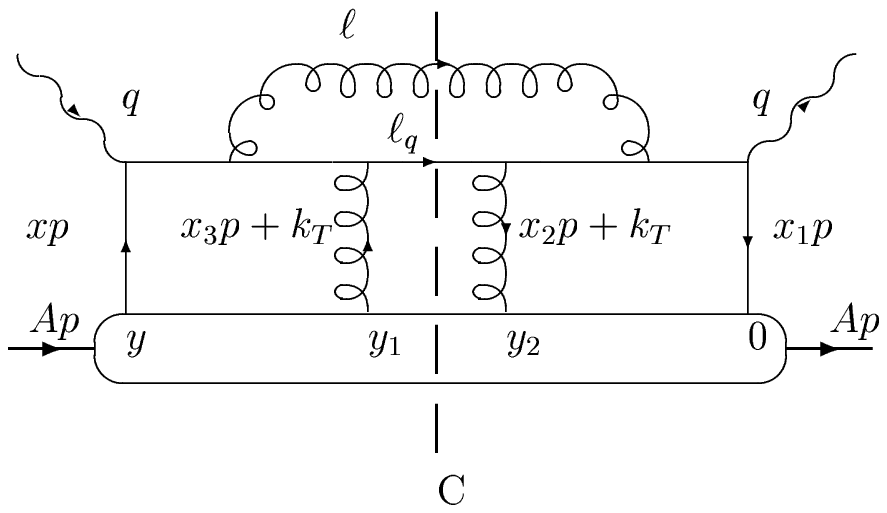,width=3in,height=2.0in}}
\caption{A central-cut diagram for quark-gluon rescattering processes.}
\label{fig1}
\end{figure}

The twist-four parton matrix elements are in principle not calculable and
can only be measured in experiments, just like twist-two parton 
distributions. However, under certain assumptions, one can
use some nuclear model to relate them to twist-two parton
distributions in nucleons.  Along the way, one obtains the $A$-dependence
of these nuclear matrix elements. 

If we assume that the nuclear wave function can be expressed as 
a multiple-nucleon state, with each nucleon a color singlet,
the two gluon fields must operate on the same nucleon state inside
the nucleus. Consider the dominant case where the quark and gluon fields
operate on different nucleons inside the nucleus. The integration over
$y_1^-$ and $y_2^-$ in Eq.~(\ref{eq:t4}) should give the length scale
$r_NR_A$, where $r_N$ is the nucleon radius and 
$R_A\simeq1.12A^{1/3}\,{\rm fm}$ is the 
radius of the nucleus. The twist-four nuclear 
parton matrix elements should then be 
approximately proportional to $A^{4/3}$. If the quark and gluon come
from the same nucleon, the matrix elements will only be proportional to
$A$, which is subleading. 

The twist-four two-parton correlation function that enters the modified
quark fragmentation function has not only normal parton matrix
elements representing direct terms in the square of the amplitude,  
but also those representing interference. The former has
momenta flowing directly along two parton fields separately while
the latter has momenta flowing across two different parton fields.
These two different contributions
were called `diagonal' and `off-diagonal' 
matrix elements, respectively, in Ref. \cite{GW}.  We 
call them `direct' and `crossed' here to avoid confusion
with truly off-diagonal matrix elements, in which the 
momenta of the external states are different.
The relative signs between
these two kinds of matrix elements reflect the physics of the LPM
interference effect in the processes of induced gluon radiation. As
illustrated by the central cut-diagram in Fig.~\ref{fig1}, the
gluon radiation can either be produced as final state radiation of
the photon-quark hard scattering or initial state radiation of the
quark-gluon rescattering. In the former case, the energy of the radiated
gluon is provided by the initial quark with $x=x_B+x_L$. The quark-gluon
rescattering can be very soft since the momentum fraction carried
by the gluon, $x_3=x_T=\langle k_T^2\rangle/2p^+q^-z$, 
is very small when
$\langle k_T^2\rangle\approx 0$. This is normally referred to as
a hard-soft process. In the latter case, however, the initial gluon 
must carry a finite momentum fraction $x_3=x_T+x_L$ 
to induce the gluon radiation.
Such a process is called double hard scattering. Contributions from
central-cut diagrams such as Fig.~\ref{fig1} contain both of these
processes as well as their interference. Since initial and final state
radiation amplitudes have a phase difference of $\pi$ , the
final result of the sum of these contributions is the dipole-like 
form-factor in the radiation spectrum
which is now absorbed into the definition of the two-parton
correlation $T_{qg}^A(x,x_T,x_L)$. As $x_Lp^+\rightarrow 0$, the effective
gluon radiation spectrum vanishes because
the interference becomes complete. This is exactly the LPM 
effect \cite{LPM}, which is now embedded in the effective two-parton
correlation function.

Because of the LPM interference effect in induced bremsstrahlung,
the effective two-parton correlation function that enters the
modified fragmentation function essentially contains four independent
twist-four nuclear parton matrix elements. The two direct
ones correspond to gluon radiation associated with 
photon-quark and quark-gluon scattering. In the corresponding forward 
scattering  processes, shown in Fig.~\ref{fig1}, momentum flows 
separately along the 
quark and gluon lines. The twist-four parton matrix elements can
then have the interpretation of a two-parton joint distribution inside 
the nucleus. The two crossed matrix elements are 
related to the inteference between the two different radiation processes. 
In this case, there is actually a momentum flow in the amount $x_Lp^+$ 
between quarks and 
gluons in the forward scattering amplitude. Such crossed matrix
elements do not have the interpretation of a normal parton distribution.
Since we only consider the case where the quark and gluon are from different
nucleons inside the nucleus, these contributions should be related to 
the off-forward (or skewed) parton distrbution functions of a nucleon. 
In the limit of vanishing skewedness ($x_L\rightarrow 0$), the crossed 
matrix elements approach the direct ones.  One can see by inspection
that the direct matrix elements are real, while the crossed ones
are complex. 

\begin{figure}
\centerline{\psfig{file=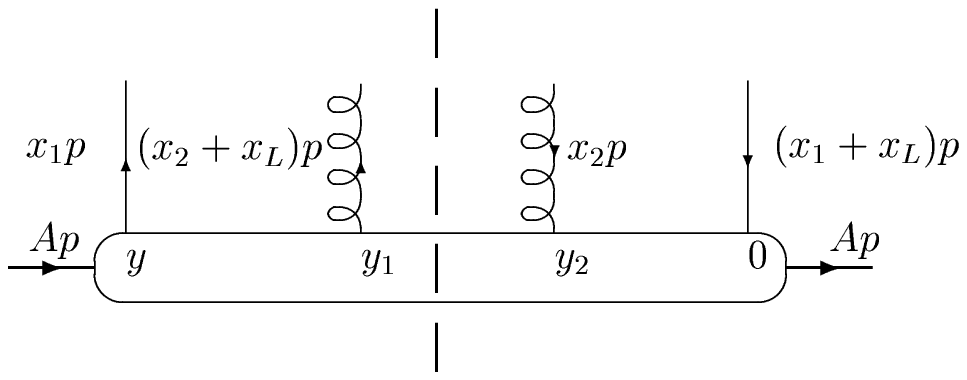,width=3.5in,height=1.5in}}
\caption{An example of a  crossed parton matrix element.  Note the 
momentum transfer from gluon to quark.}
\label{fig2}
\end{figure}

In the rest of this paper, we study these twist-four
nuclear parton matrix elements within the convolution model and relate
them to generalized nucleonic 
parton distributions. The matrix element at issue,
\begin{eqnarray}
K(x_1,x_2,x_L)&=& \int\,{dy^-\over4\pi}\,
dy^-_1\,dy^-_2\,\theta(-y^-_2)
\theta(y^--y^-_1)\nonumber\\
&&\!\!\!\!\times\;
e^{ix_1p^+y^-}\,e^{ix_2p^+(y^-_1-y^-_2)}\,
e^{ix_Lp^+y^-_1}\langle A|\overline\psi(0)\gamma^+
F_\sigma^{a\,\,+}(y^-_2)
F^{+\sigma}_a(y^-_1)\psi(y^-)|A\rangle\,\;\; ,
\label{me1}
\end{eqnarray}
describes the removal of a quark with momentum 
$x_1p$ and a gluon of momentum $(x_2+x_L)p$ from
our nuclear state $|A\rangle$, and the subsequent
replacement of a gluon with momentum $x_2p$ and 
a quark of momentum $(x_1+x_L)p$, as illustrated in Fig.~\ref{fig2}.
This nuclear parton matrix element is useful in constructing
the physical combinations that
appear in many nuclear scattering 
processes \cite{LQS}, as
well as the in-medium evolution of the parton fragmentation
functions \cite{GW}.  In particular, the correlation function
(\ref{eq:t4}) appearing in the medium-modified quark fragmentation
function (\ref{eq:dmod}) can be written as
\begin{eqnarray}
T^A_{qg}(x,x_T,x_L)&=&K(x+x_L,x_T,0)-K(x,x_T,x_L)\nonumber\\
&&-K(x+x_L,x_T+x_L,-x_L)
+K(x,x_T+x_L,0)\;\; .\label{TKsim}
\end{eqnarray}

\section{The Convolution Model}

In order to relate the matrix element (\ref{me1})
to the nucleonic degrees of freedom of the nucleus,
we must define these degrees of freedom quantitatively.
Our formalism is based on light-cone 
perturbation theory \cite{mueller},
in which one expands the physical state under consideration
in terms of the free-particle states of its constituents.
In our case, we would like to express our nuclear state in terms
of free nucleonic states.  Phenomenologically, we
know that this basis is not complete and therefore cannot
be guaranteed to span our nuclear state space. 

Contributions from higher Fock states are required
in quantum field theory to generate effective nonlocal
interactions from the underlying contact terms of the Lagrangian.
However, most of the effect of higher Fock states is already
included in the definition of the twist-two nucleonic parton
matrix elements. The additional effects of higher Fock states
in a nucleus are induced by nucleonic interactions.
However, these effects are limited by the effective 
interaction energy of the system under consideration.
Here, we are mainly concerned with nuclear systems whose
interaction energy is small compared with the energy 
per nucleon. This allows us to consider only the lowest 
Fock state, that of $A$ nucleons.  We mention here that although
we neglect interactions in the form of higher Fock
states, nucleon correlations will still occur
through the wave function in our Hilbert space.  
These correlations lead to nontrivial relations
between the nuclear matrix element and the 
nucleonic distributions.

In light of the above approximation, we write
\begin{equation}
|A\rangle \simeq \int d\Pi_A\;
\phi(\lbrace p_i\rbrace)|\lbrace p_i\rbrace\rangle
\,2p^+(2\pi)^3\delta^{(3)}\left(p_A-\sum_{i=1}^Ap_i\right)\;\; ,
\label{aapprox}
\end{equation}
where $|\lbrace p_i\rbrace\rangle$ represents the state of 
$A$ free nucleons of momenta $\lbrace p_i\rbrace$ normalized
as 
\begin{equation}
\langle\lbrace p_i'\rbrace|\lbrace p_i\rbrace\rangle
=\prod_{i}2p_i^+(2\pi)^3\delta^{(3)}(p_i-p_i')\;\; ,
\end{equation}
and 
\begin{equation}
d\Pi_A\equiv\prod_{i=1}^A\left\lbrace {d^3p_i\over(2\pi)^3}
\,{\theta(p_i^+)\over2p_i^+}\right\rbrace
\end{equation}
represents the differential phase-space of $A$ nucleons.

The nucleon states are specified by their
`+' and `$\perp$' momentum components; 
$d^3p_i\equiv dp_i^+d^2p_{i\perp}$.
The `--' component of each momentum is determined by the 
on-shell condition,
\begin{equation}
p_i^-={p_{i\perp}^2+M^2\over 2p_i^+}\;\; ,
\end{equation}
and is {\it not} conserved (i.e. $\sum_i p_i^-\neq p_A^-$).
Again, $M$ is the nucleon mass. The normalization of our nuclear state, 
$\langle A'|A\rangle=2p^+(2\pi)^3\delta^{(3)}(p_A-p_A')$ implies
\begin{equation}
\int d\Pi_A\;
|\phi(\lbrace p_i\rbrace)|^2
\,2p^+(2\pi)^3\delta^{(3)}\left(p_A-\sum_{i=1}^Ap_i\right)=1\;\; .
\label{norm}
\end{equation}
If we were to consider higher Fock states as well, 
the 1 on the right-hand side of this equation would be replaced by the 
probability of finding our state in this lowest Fock state.

Our wave function $\phi$ contains many distribution
functions.  In particular, we can define
the one-nucleon density
\begin{eqnarray}
&&\rho_1(k)\equiv\int d\Pi_{A-1}\;
|\phi(k,\lbrace p_i\rbrace)|^2\nonumber\\
&&\qquad\times 2p^+(2\pi)^3\delta
\left(Ap^+ -k^+-\sum_{i=1}^{A-1}p_i^+\right)
\delta^{(2)}\left(k_\perp+\sum_{i=1}^{A-1}p_{i\perp}\right)\;\; ,
\end{eqnarray}
which represents the probability of finding
a nucleon of momentum $k$ in our nucleus 
irrespective of the momenta of the other nucleons.
In light of Eq.~(\ref{norm}), $\rho_1$ satisfies
\begin{equation}
\int {d^3k\over(2\pi)^3}\,{\theta(k^+)\over 2k^+}\,
\rho_1(k)=1\;\; .
\end{equation}
Hence the light-cone nucleon distribution 
function in Eq.(\ref{conv}) can be written as
\begin{equation}
\rho(\alpha)=\frac{1}{4\pi \alpha}
\int{d^2k_\perp\over(2\pi)^2}\rho_1(k)\;\; ,
\end{equation}
where $\alpha\equiv k^+/p^+$.
Eq.(\ref{conv}) comes directly from the substitution 
of (\ref{aapprox}) into the definition of the nuclear quark 
distribution function $f_{q/A}(x)$ in Eq.~(\ref{eq:tw2}).

The two-nucleon correlator, 
\begin{eqnarray}
&&\rho_2(k_1;k_{2};\Delta)\equiv\int d\Pi_{A-2}\;
\phi^*(k_{1}-\Delta/2,
k_{2}+\Delta/2,\lbrace p_i\rbrace)
\phi(k_{1}+\Delta/2,k_{2}
-\Delta/2,\lbrace p_i\rbrace)\nonumber\\
&&\qquad\times\;2p^+(2\pi)^3
\delta\left(Ap^+-k_1^+-k_2^+ -\sum_{i=1}^{A-2}p_i^+\right)
\delta^{(2)}\left(k_{1\perp}+ k_{2\perp}
+\sum_{i=1}^{A-2}p_{i\perp}\right)\;\; ,
\end{eqnarray}
contains information about the {\it sharing} of
momentum by nucleons inside the nucleus and 
appears in the double scattering process we consider.
The two-nucleon density, 
$\rho_2(k_1;k_2;0)$,
represents the probability of finding 
two nucleons with the specified momenta 
within the nucleus, and is normalized as
\begin{equation}
\int d\Pi_2\;\rho_2(k_1;k_2;0)=1\;\; .
\label{rho2norm}
\end{equation}
The use of this
function rather than the two-parton correlator, $\rho_2$, leads to the 
expectation of diagonal parton distributions 
in twist-four nuclear matrix elements.  

Due to the $\theta$-functions, our matrix element cannot
readily be interpreted as a product of twist-two distributions 
as it stands.  These $\theta$-functions order the 
fields along the light-cone axis to make the multiple
scattering process physical.
Employing the representation
\begin{equation}
\theta(y)=\mp{1\over2\pi i}\int\,dz\,{1\over z-x\pm i\varepsilon}
e^{-i(z-x)y}
\end{equation}
for the $\theta$-function, our nonperturbative
distribution takes the form
\begin{eqnarray}
K(x_1,x_2,x_L)&=&{1\over2}\int{dz_1\over z_1-\omega_1-i\varepsilon}\;
{dz_2\over z_2-\omega_2+i\varepsilon}\;\nonumber\\
\label{element}
&&\times\int{dy^-\over2\pi}\,{dy^-_1\over2\pi}
\,{dy^-_2\over2\pi}
e^{ip^+y^-(x_1+z_1-\omega_1)}e^{ip^+y^-_1(x_2+x_L-z_1+\omega_1)}
e^{ip^+y^-_2(z_2-\omega_2-x_2)}\\
&&\qquad\times\left\langle A\left|
\overline\psi(0)\gamma^+\psi(y^-)
F^{a+}_\sigma(y^-_2)F^{+\sigma}_a
(y^-_1 )\right|A\right\rangle\;\; .\nonumber
\end{eqnarray}
Here, $\omega_1$ and $\omega_2$ are arbitrary real variables that 
will be chosen later to simplify the final results. 

Substitution of our approximate nuclear state
(\ref{aapprox}) into (\ref{element}) leads to matrix elements
of the form
\begin{equation}
\langle \lbrace p_i'\rbrace|\overline\psi(0)\gamma^+\psi(y^-)\;
F^{a\;+}_\sigma(y^-_2)F_a^{+\sigma}(y^-_1)
|\lbrace p_i\rbrace\rangle\;\; .
\end{equation}
Assuming that the color correlation length along the light-cone
within our nucleus is not larger than the nucleon size and
neglecting the effects of direct multi-nucleonic correlations
as higher twist,
we can factorize this expression into a product
of single-particle Hilbert space amplitudes:
\begin{eqnarray}
&&A\langle p_1'|\overline\psi(0)\gamma^+\psi(y^-)\;
F^{a\;+}_\sigma(y^-_2)F_a^{+\sigma}(y^-_1)
|p_1\rangle\,\prod_{i=2}^A 2p_i^+(2\pi)^3\delta^3(p_i'-p_i)\nonumber\\
&&+{A\choose2}
\langle p_1'|\overline\psi(0)\gamma^+\psi(y^-)|p_1\rangle\;
\langle p_2'|F^{a\;+}_\sigma(y^-_2)F_a^{+\sigma}(y^-_1)
|p_2\rangle \;\prod_{i=3}^A 2p_i^+(2\pi)^3\delta^3(p_i'-p_i)\;\; .
\label{eq:decomp}
\end{eqnarray}
Our dismissal of direct multi-nucleonic correlations 
is one of the main approximations in this paper, and 
will be discussed in more detail in the conclusions.
For now, we note that these higher-twist corrections are suppressed
by powers of $Q^2$ and as such can be neglected at
large scales.

The contributions to (\ref{eq:decomp}) can be interpreted 
in terms of the multiple parton scattering picture in 
DIS.  Matrix elements related to 
$\overline\psi(0)\gamma^+\psi(y^-)$
represent the probability that a quark in a certain
nucleon is struck by our probe.
The struck quark then propagates through the nucleus, 
encountering  another parton at some point during its 
journey.  If this rescattering 
occurs while the struck parton is still in its 
parent nucleon, the effect is represented by a twist-four nucleonic 
matrix element convoluted with the single-nucleon density. This
is essentially the first term in Eq.(\ref{eq:decomp}).
Considering the coherence length of the scattering to be of the order
of the nucleon size, this term should be proportional to $A$.

If the rescattering is with a parton in another nucleon, the
probability is related to the twist-two parton distributions
in each nucleon convoluted with the two-nucleon correlator.
We can call this double-factorized rescattering.
Since the nuclear radius grows with $A^{1/3}$, one expects
this double scattering process to be proportional to $A^{4/3}$,
enhanced by $A^{1/3}$ over the double scattering within the same
nucleon. The first term in the above decomposition is therefore
suppressed relative to the second as $A$ increases.

Using the above decomposition, and keeping only 
the contribution of the 
double-factorized rescattering, $K(x_1,x_2,x_L)$ becomes
\begin{eqnarray}
&&K(x_1,x_2,x_L)\simeq{2}
{A\choose2}\int{dz_1\over z_1-\omega_1-i\varepsilon}
{dz_2\over z_2-\omega_2+i\varepsilon}
\int{d^2k_{1\perp}\over(2\pi)^2}
{d^2k_{2\perp}\over(2\pi)^2}
{d^2\Delta_{\perp}\over(2\pi)^2}\nonumber\\
&&\qquad\times\int
{d\alpha_1\over2\pi}\,
{d\alpha_2\over2\pi}\,
{d\zeta\over2\pi}\,
{\theta(\alpha_1+\zeta)\over2(\alpha_1+\zeta)}
\,{\theta(\alpha_1-\zeta)\over2(\alpha_1-\zeta)}\,
{\theta(\alpha_2+\zeta)\over 2(\alpha_2+\zeta)}\,
{\theta(\alpha_2-\zeta)\over 2(\alpha_2-\zeta)}\nonumber\\
&&\qquad\qquad\times\;\delta(x_L-2\zeta+z_2-z_1-\omega_2+\omega_1)
\rho_2(k_{1};k_{2};\Delta)\\
&&\qquad\qquad\times\;\int{dy^-\over2\pi}
e^{ip^+y^-(x_1+z_1-\omega_1+\zeta)}\left\langle k_1-{\Delta\over2}\left|
\overline\psi\left(-{y^-\over2}\right)\gamma^+
\psi\left({y^-\over2}\right)
\right|k_1+{\Delta\over2}\right\rangle\nonumber\\
&&\qquad\times\;\int{dy^-_d\over2\pi}\,{1\over p^+}
e^{ip^+y^-_d(x_2+x_L-z_1+\omega_1-\zeta)}
\left\langle k_2
+{\Delta\over2}\left|F^{a\;+}_\sigma\left(-{y^-_d\over2}\right)
F_a^{+\sigma}\left({y^-_d\over2} \right)
\right|k_2-{\Delta\over2}\right\rangle\;\; ,
\nonumber
\end{eqnarray}
where $y_d=y_1-y_2$, $\alpha_i\equiv k_i^+/p^+$ 
and $\zeta\equiv-\Delta^+/2p^+$.

If we could take $\Delta=0$, this expression 
would be reduced to the ordinary parton distributions
probed in deeply-inelastic scattering.  
As it is, we can employ the off-forward parton distributions
(OFPD's) \cite{ji} [Note that the distributions used here
are slightly different than those defined in \cite{ji}.
In particular, $G(x,\xi,t)=xF_g(x,\xi,t)$].  Thus,
\begin{eqnarray}
\int{dy^-\over2\pi}\,e^{ip^+y^- x}\left\langle 
k_1-{\Delta\over2}\left|\overline\psi\left(-{y^-\over2}\right)
\gamma^+\psi\left({y^-\over2}\right)
\right|k_1+{\Delta\over2}\right\rangle&=&2
Q\left({x\over \alpha_1},-\zeta,{t_1\over M^2}\right)\;\; ,\\
\int{dy^-\over2\pi}\,e^{ip^+y^- x}\left\langle 
k_2+{\Delta\over2}\left|
F^{a\;+}_\sigma\left(-{y^-\over2}\right)
F_a^{+\sigma}\left({y^-\over2}\right)\right|k_2
-{\Delta\over2}\right\rangle
&=&2\alpha_2\,p^+G\left({x\over \alpha_2},{\zeta},{t_2\over M^2}\right)\;\; ,
\end{eqnarray}
where 
\begin{equation}
t_i=-{4\zeta^2M^2\over \alpha_i^2-\zeta^2}-{(2\zeta \vec p_{i\perp}
+\alpha_i\vec \Delta_\perp)^2\over \alpha_i^2-\zeta^2}
\end{equation}
represents the squared four-momentum 
transfers for the two matrix elements.
Note $t_1\neq t_2$ since the `-'-components of our four-vectors
are not conserved in this version of perturbation theory.

In terms of these OFPD's, we have
\begin{eqnarray}
K(x_1,x_2,x_L)&\simeq&
{1\over2}{A\choose2}\int
{d\alpha_1\over2\pi}\,
{d\alpha_2\over2\pi}\,
{d\zeta\over2\pi}\,
{\theta(\alpha_1+\zeta)\over \alpha_1+\zeta}\,
{\theta(\alpha_1-\zeta)\over \alpha_1-\zeta}\,
{\theta(\alpha_2+\zeta)\over \alpha_2+\zeta}\,
{\theta(\alpha_2-\zeta)\over \alpha_2-\zeta}
\nonumber\\
\label{mef}
&&\times\int{d^2k_{1\perp}\over(2\pi)^2}
{d^2k_{2\perp}\over(2\pi)^2}
{d^2\Delta_{\perp}\over(2\pi)^2}
\rho_2(k_1;k_{2};\Delta)\\
&&\times\int{dz_1\over (z_1-\omega_1-i\varepsilon)}
\;{dz_2\over (z_2-\omega_2+i\varepsilon)}
\delta(x_L-2\zeta+z_2-z_1-\omega_2+\omega_1)\nonumber\\
&&\!\!\!\!\!\!\!\!\times\; \alpha_2 
Q\left({x_1+z_1-\omega_1+\zeta\over \alpha_1}
,-{\zeta},{t_1\over M^2}\right)
G\left({x_2+x_L-z_1+\omega_1-\zeta\over \alpha_2},
{\zeta},{t_2\over M^2}\right)\;\; .
\nonumber
\end{eqnarray}

This complicated expression can be reduced to a more 
enlightening form through a few plausible assumptions on the
form of our nucleonic correlation $\rho_2$ and the 
OFPD's.  To begin with, we assume that $\rho_2$
is peaked around $\alpha_i=1$ and $\zeta=\Delta_\perp=k_{i\perp}=0$,
with widths that are governed by the nuclear radius, $R_A$.
This ansatz is dictated by the expectation that
the nucleons are confined within the nuclear
radius in position space.
Specifically, we write 
\begin{equation}
\rho_2(\alpha_1,k_{1\perp};\alpha_2,k_{2\perp};-2\zeta,\Delta_\perp)
= R_A^4/x_A^2\; r_2\left({\alpha_1-1\over x_A},
R_Ak_{1\perp};
{\alpha_2-1\over x_A},R_Ak_{2\perp};
{-2\zeta\over x_A},R_A\Delta_\perp\right)\;\; ,\label{r2def}
\end{equation}
with $x_A=1/(MR_A)$ 
and $r_2$ approximately independent
of $A$.  The behavior of the normalization $R_A^4/x_A^2$
as a function of $A$ can be determined
from the normalization condition (\ref{rho2norm}) on our 
two-particle density.  Using the $\delta$-function
to perform the $\zeta$ integration and making the changes
\begin{eqnarray}
\nu_i&=&(\alpha_i-1)/x_A\\
\vec v_{i\perp}&=&R_A\vec k_{i\perp}\\
\vec\delta_\perp&=&R_A\vec\Delta_\perp\\
z&=&(z_1+z_2)/2x_A\\
u&=&(z_1-z_2)/2x_A
\end{eqnarray}
in variables, we arrive at the expression
\begin{eqnarray}
K(x_1,x_2,x_L)&\simeq&
{1\over4\pi R_A^2}{A\choose2}
\int du\,dz {1\over z+u+\xi-i\varepsilon}\;
{1\over z-u-\xi+i\varepsilon}\nonumber\\
&&\times\;\int{d\nu_1\over2\pi}\,{d\nu_2\over2\pi}\,
{\theta(1+x_A(\nu_1-u))\over 1+x_A(\nu_1-u)}\,
{\theta(1+x_A(\nu_1+u))\over 1+x_A(\nu_1+u)}\, \nonumber \\
&& \times
{\theta(1+x_A(\nu_2-u))\over 1+x_A(\nu_2-u)}\,
{\theta(1+x_A(\nu_2+u))\over 1+x_A(\nu_2+u)}\,
\nonumber\\
\label{me2}
&&\times\int{d^2v_{1\perp}\over(2\pi)^2}
{d^2v_{2\perp}\over(2\pi)^2}
{d^2\delta_{\perp}\over(2\pi)^2}
r_2(\nu_1,\vec v_{1\perp};\nu_2,\vec v_{2\perp};2u,\vec\delta_\perp)
(1+x_A\nu_2) \\
&&\times\; Q\left({x_1+x_A(\xi+z)\over 1+x_A\nu_1},
{x_Au},{\tilde t_1}\right)
G\left({x_2+x_A(\xi-z)\over1+x_A \nu_2},
-{x_Au},{\tilde t_2}\right)\;\; ,\nonumber
\end{eqnarray}
for (\ref{mef}).  For simplicity, we have 
chosen $\omega_1=-\omega_2=-x_L/2$ and defined the  
parameters $\xi\equiv x_L/(2x_A)$ and  
\begin{equation}
\tilde t_i \equiv-{x_A^2\over (1+x_A \nu_i)^2-x_A^2 u^2}
\left\lbrack4u^2+((1+x_A\nu_i)\vec\delta_\perp-
2x_Au\vec v_{i\perp})^2\right\rbrack\;\; .
\end{equation}

Working from this form of $K(x_1,x_2,x_L)$,
it is easy to see the large-$A$ enhancement
of the multiple-scattering contribution.
Since $r_2$ depends only weakly on $A$, 
this contribution to $K$ scales like 
$A^{4/3}$ as expected.  We can simplify our
expression further by assuming that $r_2$ 
is {\it sharply} peaked in the sense that
all of its moments are finite.  In this
case, we can expand
our integrand about the peak in $r_2$
in the formal limit $A\rightarrow\infty$
and drop all non-leading terms :
\begin{eqnarray}
K(x_1,x_2,x_L)&\simeq&
{A^2\over8\pi R_A^2}
\int du\,dz {1\over z+u+\xi-i\varepsilon}\;
{1\over z-u-\xi+i\varepsilon}
\tilde r_2(u) \nonumber\\
&&\times\;Q(x_1+x_A(\xi+z),x_A u,0)G(x_2+x_A(\xi-z),-x_Au,0)\;\; ,
\label{simple}
\end{eqnarray}
where 
\begin{equation}
\tilde r_2(u)\equiv
\int{d\nu_1\over2\pi}\,{d\nu_2\over2\pi}\,
{d^2v_{1\perp}\over(2\pi)^2}\,
{d^2v_{2\perp}\over(2\pi)^2}\,
{d^2\delta_{\perp}\over(2\pi)^2}
r_2(\nu_1,v_{1\perp};\nu_2,v_{2\perp};2u,\delta_\perp)\;\; .
\end{equation}

Expression (\ref{simple}) is the main result 
of this paper.  Its derivation requires only
the assumptions that the lowest Fock state dominates 
the nuclear wave function and that the nucleonic correlator
is sharply peaked with a width dictated by the 
nuclear radius.  Strictly speaking, 
this expression is valid only
in the formal limit $A\rightarrow \infty$.
For any finite value of $A$, one must investigate 
the size of the derivatives of the OFPD's in 
relation to $A^{1/3}$.  While this investigation
must be done within a specific model, the
contributions are expected to be small
as long as the singular regions are avoided.
Our implicit assumption that the
OFPD's are analytic functions
of the virtuality of the momentum transfer, $t_i$, is supported
by studies of these functions \cite{ji,radush}.  
On the other hand, the kernels 
dictating the evolution of these functions (cf \cite{ji2})
imply that the OFPD's are {\it not} analytic
functions of their second argument.
This is why we have not expanded our integrand 
about $x_Au=0$.  

To put this expression into a form suitable for 
numerical evaluation, we write 
\begin{equation}
{1\over z+u+\xi-i\varepsilon}\;\times\;{1\over z-u-\xi+i\varepsilon}
={\rm P}\left\lbrack\vphantom{{1\over z+u+\xi-i\varepsilon}}
{1\over 2z}\right\rbrack\left\lbrack
{1\over z+u+\xi-i\varepsilon}+{1\over z-u-\xi+i\varepsilon}\right\rbrack
\end{equation}
and explicitly separate the $u$-integration into its pole and principal
value parts.  After a change of variables, this leads to
\begin{eqnarray}
K(x_1,x_2,x_L)\simeq{A^2\over 8\pi R_A^2}\left\lbrace
\int_0^\infty{dz\over 2z}\,\int_0^\infty{dw\over w}
\left\lbrack F(z,w-z-\xi)
-F(z,w+z+\xi)\right.\right.\nonumber\\
\left.+F(z,w-z+\xi)
-F(z,w+z-\xi)\right\rbrack\nonumber\\
\left.
+i\pi\int_0^\infty{dz\over 2z}
\left\lbrack F(z,z+\xi)-F(z,z-\xi)\right\rbrack\right\rbrace\;\; ,
\label{explicit}
\end{eqnarray}
where 
\begin{eqnarray}
F(z,u)\equiv \left\lbrack Q(x_1+x_L/2+x_Az,x_Au,0)G(x_2+x_L/2-x_Az,x_Au,0)
\right.\nonumber\\
\left.+Q(x_1+x_L/2-x_Az,x_Au,0)G(x_2+x_L/2+x_Az,x_Au,0)\right\rbrack
\tilde r_2(u)\;\; .
\end{eqnarray}
In deriving this result, we have used the facts that
$\tilde r_2(u)=\tilde r_2(-u)$ and that the OFPD's are even functions
of their second argument, as can be seen by inspection.

Several features of this expression are worth pointing out.
First, we note that the imaginary part of our matrix element is
odd in $\xi$.  This causes the combination in Eq.(\ref{TKsim}) 
to be real, as required.  One can check that the two-parton correlation
$T^A_{qg}(x,x_T,x_L)$, as expressed in Eq.~(\ref{TKsim}), depends
only on the real part of $K(x_1,x_2,x_L)$.
In addition, this
combination is necessarily positive.  The fact that these consistency
requirements are satisfied is gratifying, but not unexpected.
On the other hand, the relationship between our expression and the naive
expectation 
\begin{equation}
K(x_1,x_2,0)\sim Q(x_1)G(x_2)\;\; ,
\label{naiveexp}
\end{equation}
where $Q(x)\equiv Q(x_1,0,0)$ and $G(x)\equiv G(x,0,0)$ are the 
ordinary diagonal distributions, 
is entirely unclear.  Ignoring, for the moment, the non-analytic 
nature of the OFPD's, one can imagine expanding each term in the 
integrand about the peak of the nuclear sharing function $\tilde r_2$ and
dropping all higher-order terms.  This reduces the 
OFPD's to ordinary parton distribution functions, bringing
us closer to (\ref{naiveexp}).  However, we cannot
formally reproduce this simple dependence because of a
remaining convolution between the distributions.
This convolution causes our matrix element to sample the 
entire parton distribution functions, regardless of the values
of $x_1$, $x_2$, and $x_L$.  Its presence is a direct consequence
of the correlative nature of the matrix element.  Mathematically,
it comes from the $\theta$-functions.  

We could attempt to 
remove the correlation by expanding our result about $x_A=0$. 
However, corrections to the leading term in our expansion
diverge, indicating
nonanalyticity at $x_A=0$.  This does not mean that the 
leading term is not a good approximation to the 
full solution, only that the corrections cannot 
be expressed in terms of powers of $x_A$.  
Nevertheless, assuming that the corrections are small 
for some range of $x_A$ near zero, one can compute 
Eq.~(\ref{simple}) via contour integration directly
in the limit of $x_A\rightarrow 0$, but with fixed value
of $\xi=x_L/2x_A$. The result,
\begin{eqnarray}
K(x_1,x_2,x_L)\left.\vphantom{x^5\over x^6}\right|_{x_A\rightarrow 0}
&\equiv& K_0(x_1,x_2,x_L) \nonumber \\
&\simeq& {\pi A^2\over 8 R_A^2}\;Q(x_1+x_L/2)G(x_2+x_L/2) \nonumber \\
&\times&\left\{\tilde r_2(\xi) + \frac{i}{\pi}\int_0^{\infty}\frac{du}{u}
\left[\tilde r_2(u+\xi)-\tilde r_2(u-\xi)\right]\right\}
\;\; ,\label{expect}
\end{eqnarray}
is of the form of the naive expectation.  
In order to determine the validity of this
approximation, we must calculate $K$ numerically
in some model and compare the results to the above approximation.

\section{Numerical Results}

In this section, we explore some of the properties
of expression (\ref{simple}) in a specific model.
While this model is certainly not expected
to conform to the quantitative details of
the realistic nuclear wave functions, 
we expect its general features to
be echoed in more realistic treatments.

For simplicity, we assume a Gaussian
form for $\tilde r_2$ :
\begin{equation}
\tilde r_2(u)={\langle R_A^2\Delta_\perp^2\rangle\over4\pi}e^{-u^2}\;\; .
\end{equation}
The normalization is determined via (\ref{rho2norm}) 
in conjunction with (\ref{r2def}) by assuming a Gaussian 
dependence of $r_2$ on $\delta_\perp$.  The constant
$\langle R_A^2\Delta_\perp^2\rangle$ is a measure of the 
transverse momentum-sharing among nucleons in our nucleus, and
is expected to be of order one.
Expanding our integrands about the peak in $\tilde r_2$ and performing
the integral over $u$, we arrive
at the expression
\begin{eqnarray}
K(x_1,x_2,x_L)&\simeq& 
{A^2\over 32\pi^{3/2} R_A^2}\;\langle R_A^2\Delta_\perp^2\rangle
\int_0^\infty{dz\over z}f(z,0) \nonumber \\
&\times& \left\lbrace \left\lbrack D(z+\xi)
+D(z-\xi)\right\rbrack +\frac{i\sqrt{\pi}}{2}
\left\lbrack e^{-(z+\xi)^2}-e^{-(z-\xi)^2}\right\rbrack
\right\rbrace\;\; .\label{modform}
\end{eqnarray}
Here, 
\begin{eqnarray}
f(z,0)\equiv Q(x_1+x_L/2+x_Az)G(x_2+x_L/2-x_Az)\nonumber\\
+Q(x_1+x_L/2-x_Az)G(x_2+x_L/2+x_Az)
\end{eqnarray}
and 
\begin{equation}
D(x)\equiv \int_0^x\;dt\; e^{\,t^2-x^2}
\end{equation}
is Dawson's integral.  Since $D(x)\rightarrow 1/2x$ as 
$x\rightarrow\infty$, $K(x_1,x_2,x_L)$ is obviously non-analytic
in $x_A$ at $x_A=0$, as mentioned above.  However, 
it can be checked explicitly that this expression reduces 
to (\ref{expect}) when $x_A\rightarrow 0$.  

In deriving Eq.~(\ref{modform}), we have assumed $x_Au\ll 1$
in the expansion around the peak of $\tilde r_2(u)$. Such an
approximation does not necessarily represent the leading behavior
of the integral.  The pole contributions occur in the 
region $u\sim \xi$, where $x_Au \sim x_L/2$ can in fact
be of the same order or larger than the first argument
of our OFPD's,  
\begin{equation}
Q(x_1+x_L/2\pm x_Az, x_L/2,0)G(x_2+x_L/2\mp x_Az, x_L/2,0).
\end{equation}
In this case, the full generalized distributions are sampled.
However, in the relevant region of the $z$-integration in
Eq.~(\ref{explicit}) with a Gaussian form of $\tilde r_2(u)$, 
$|z|\lsim\xi$. This causes the first variables in the
OFPD's to be bounded by $x_1$ and $x_2$, respectively.
According to model studies of the OFPD's \cite{radush,ofpd}, the
OFPD's can be approximated by the ordinary
parton distributions when the first argument is 
larger in magnitude than the second.  In addition, 
since the OFPD's are continuous and $G(x,\xi,t)$
is expected to be positive definite, the variation 
of $G(x,\xi,t)$ is small in the region $x<|\xi|\ll 1$.
Hence we expect very small
deviations from the ordinary gluon momentum distribution
even in the region $x_2<\!\!<x_L<\!\!<1$.  
The singular nature of the quark distribution
function $Q(x)$ at $x=0$ leads to large variations 
of the associated OFPD in the region
$x_1<x_L$.  This will essentially be
the limitation of our approximation here. 
Outside of our applicability region, the full generalized
parton distributions are needed to predict
the behavior of the matrix element.

\begin{figure}
\centerline{\psfig{file=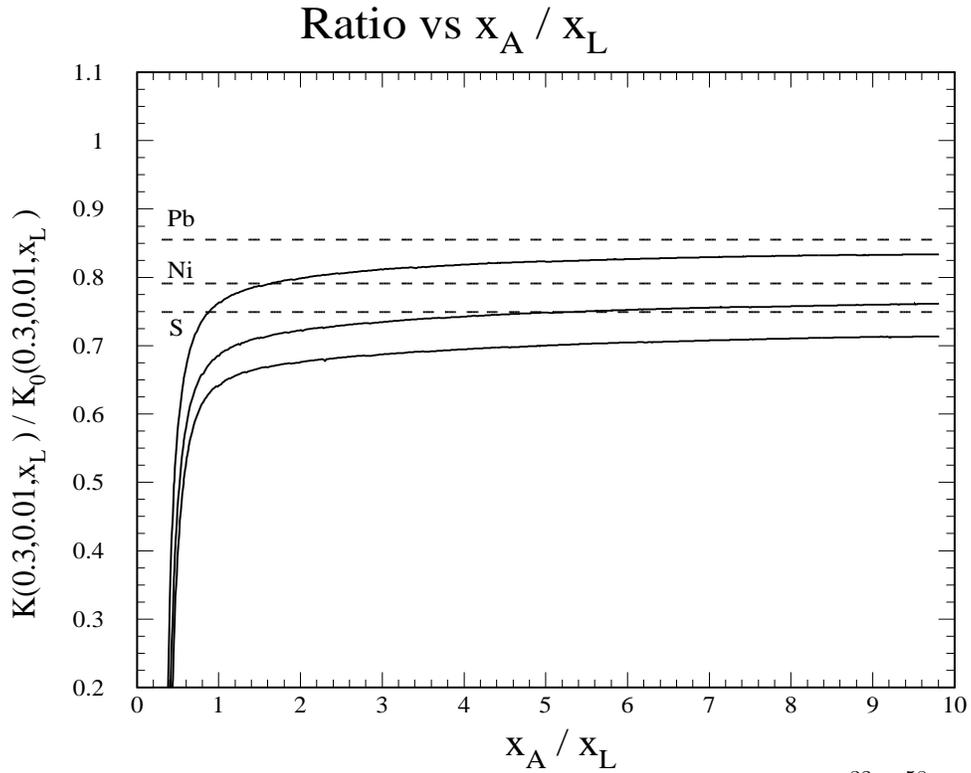,width=5.0in,height=4.0in}}
\caption{The dependence of the ratio of $K$ to $K_0$ 
on $x_L$ shown for the nuclei $^{32}$S, $^{58}$Ni, and $^{208}$Pb.  
The dashed lines show the saturation ratio for each nucleus
($x_L\rightarrow0$).  We note that although the 
saturation ratios are not very close to 1, the curves are
quite flat when $x_A>x_L$, and the ratio increases with nuclear size.}
\label{xlfig}
\end{figure}

Using the CTEQ parameterization of parton distributions  
from data \cite{cteq}, we can
calculate our expression numerically and see how 
well the results follow the factorized form $K_0$.
Since only the real parts of
$K$ enters the two-parton correlation $T^A_{qg}$, we will
concentrate on the real part of our matrix element.
The nature of (\ref{modform}) is such that 
our matrix element samples the parton momentum 
distribution at all values of $x$ rather than 
just those close to $x_{1,2}$. Since the parton distributions
are not known for $x\rightarrow 0$, we assume a simple
extrapolation with a constant value for gluons and set the
quark distributions to zero beyond the region of parametrization.
The errors introduced by such extrapolation are negligible since
the contributions to the integral from this region is very small.

Figure \ref{xlfig} shows the ratio $K(0.3,0.01,x_L)/K_0(0.3,0.01,x_L)$ 
as a function of $x_A/x_L$ for three different values of 
$x_A$.  Since $x_L$ measures the momentum sharing between 
nucleons in our matrix element, it should be smaller than the 
characteristic momentum fraction, $x_A$, in our nucleus.  
If $x_L$ becomes too large, the momentum transfer is
suppressed beyond the simple exponential suppression in $K_0$.  
This explains the 
behavior of the curve for small $x_A/x_L$.  As $x_A/x_L$ increases,
the ratio approaches the direct contribution represented
by the dashed line in the figure. 
As can easily be seen from the graph, there is 
some residual dependence on the nuclear size.  However, the
saturation ratio changes by less than 15\% as one 
changes $A$ from 208 to 32.  

\begin{figure}
\centerline{\psfig{file=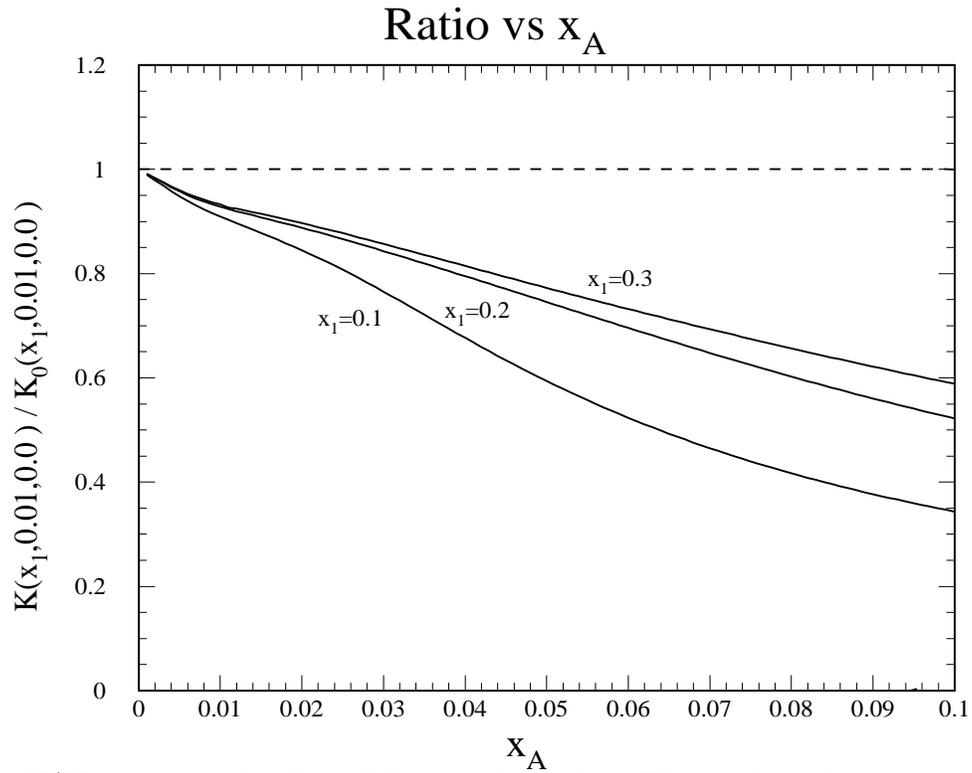,width=5.0in,height=4.0in}}
\caption{$K/K_0$ versus $x_A$ for three different values
of $x_1$.  The $x_A$ dependence is approximately
linear for small $x_A$, with a slope whose magnitude
decreases as $x_1$ increases.  Although this dependence
is nominally of order $x_A$, it can lead to quite
large corrections for real nuclei if $x_1$ is too small.}
\label{xafig}
\end{figure}

To study the residual dependence of our
matrix element on the nuclear size, we plot
the ratio $K/K_0$ at $x_L=0$ as a function of $x_A$ for three
different values of $x_1$ in Figure \ref{xafig}.
The ratio is seen to drop monotonically as $x_A$ is
increased, but the close proximity of the curves
$x_1=0.2$ and $x_1=0.3$ indicates only slight dependence on
$x_1$ when $x_1$ is moderate.  As $x_1$ becomes smaller,
the dependence on both $x_1$ and $x_A$ becomes far more dramatic.
In either case, the $x_A$ dependence is approximately linear.
Since we have already dropped many terms of this order
in $x_A$, this behavior is not at all surprising.  However,
it can lead to large corrections to the factorized form 
for real nuclei (where $x_A$ is of order 0.04).

\begin{figure}
\centerline{\psfig{file=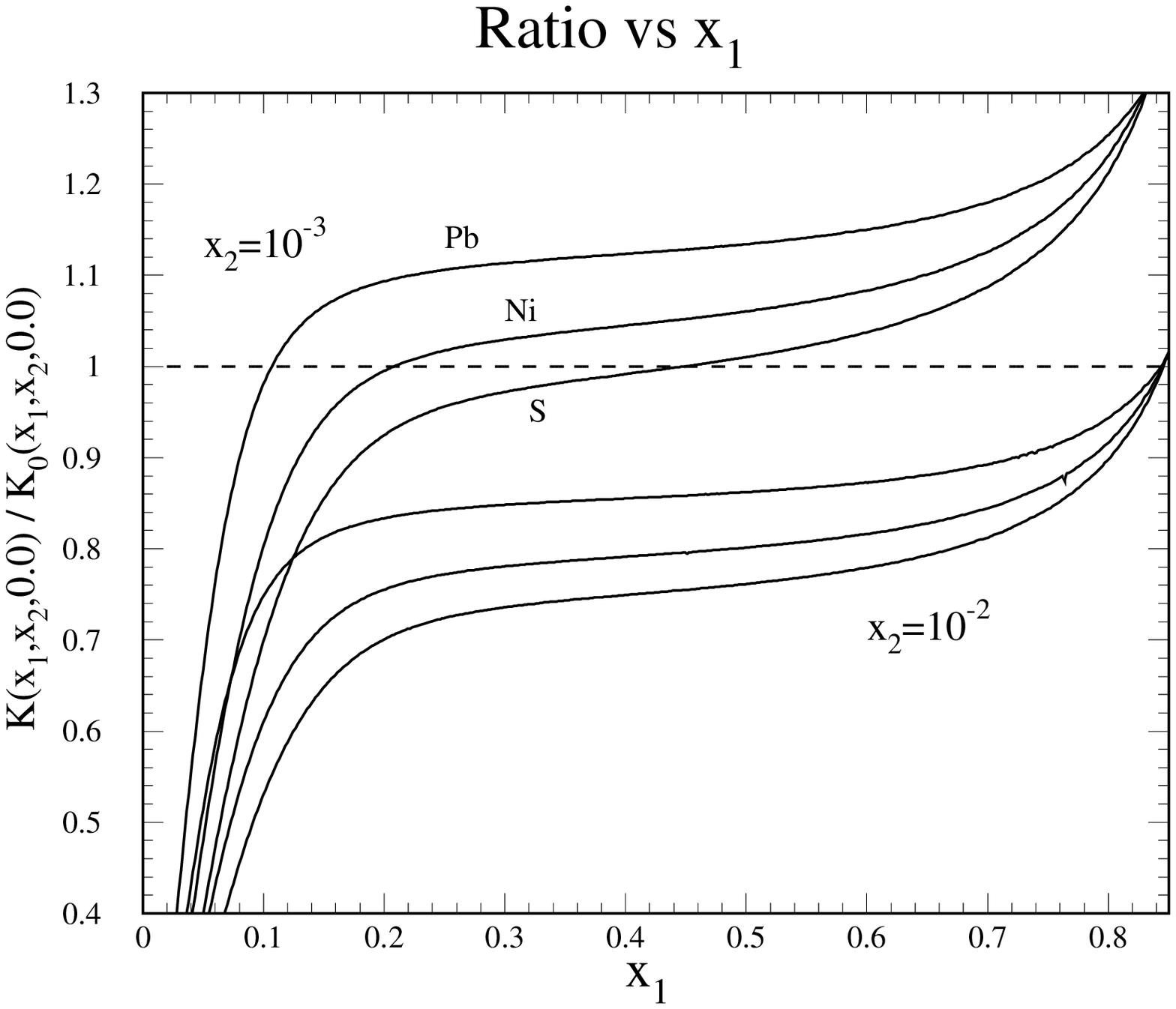,width=5.0in,height=4.0in}}
\caption{$K/K_0$ versus $x_1$ for $^{32}$S, $^{58}$Ni, and $^{208}$Pb.  
Although there is strong dependence, the ratio is 
quite flat for moderate $x_1$.  The strong dependence of 
the placement of these curves on the value of $x_2$
is attributed to the decrease of $G(x)$ for extremely 
small values of $x$.}
\label{x1fig}
\end{figure}

The $x_1$ dependence is clearly illustrated in Figure \ref{x1fig}.
Here, we see that the factorized form $K_0$ is indeed a good 
approximation for moderate $x_1$.  In the region 
$0.15\lsim x_1\lsim 0.75$, the ratio changes by approximately 15\%
for $^{208}$Pb,
while in the more restricted region $0.2\lsim x_1\lsim 0.7$, 
the change is less than 8\%.  For smaller nuclei, the 
region of `moderate $x_1$' is more restricted.
The behavior of our curves for
small $x_1$ is due to the fact that the denominator 
of our ratio diverges as $x_1\rightarrow0$, while the 
numerator samples the distribution function and smears the 
divergence.  For large $x_1$, the behavior is 
quite similar to that of the ordinary convolution model, 
Eq.(\ref{conv}), which is due to the fact that the denominator
vanishes as $x_1\rightarrow 1$ while the numerator remains finite
because the partons share momentum.
This effect is usually referred to as Fermi motion.
The dependence of the ratio on nuclear size is not too large 
(on the order of 15\% for moderate $x_1$) in the range 
$0.032\leq x_A\leq0.059$ considered.
The tremendous dependence of the placement of these curves
on the value of $x_2$ can be attributed to the turnover of the 
gluon momentum distribution at small $x$.  Since this turnover occurs
at small $x$, we expect large variation only when $x_2$ is extremely
small.

\begin{figure}
\centerline{\psfig{file=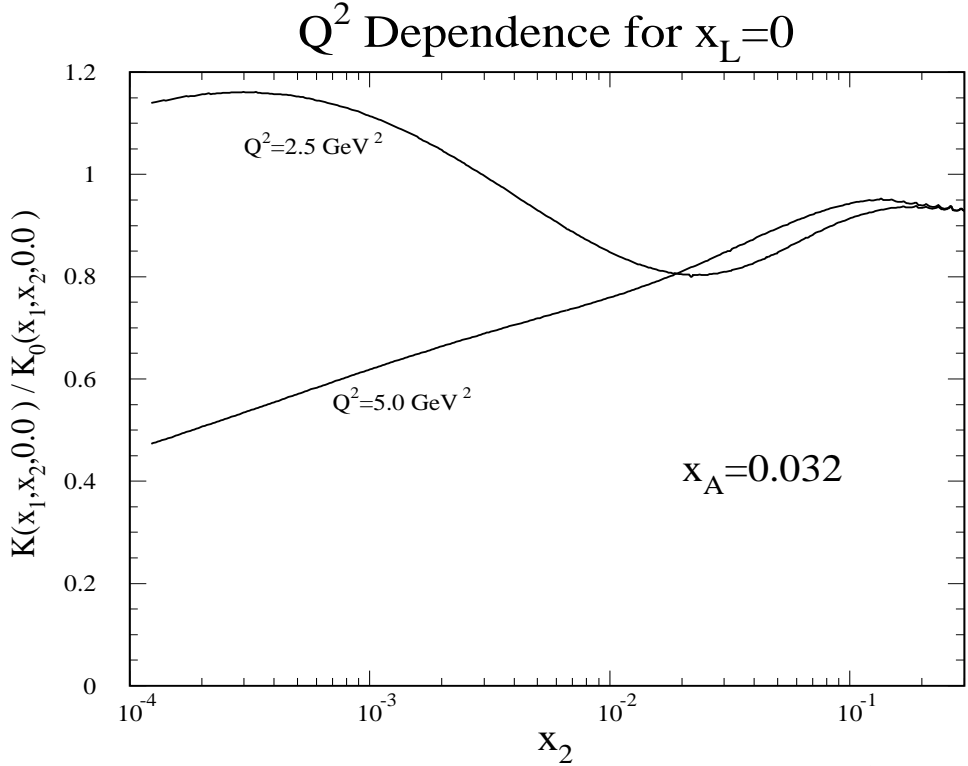,width=5.0in,height=4.0in}}
\caption{The dependence of our ratio on $x_2$
for two different values of $Q^2$.  The increase 
in the ratio as $x_2$ decreases at 2.5 GeV$^2$
is attributed to the turnover of the input gluon momentum
distribution.  Since this turnover is pushed back beyond our
cut-off at $x_2=10^{-4}$ for $Q^2=5.0$ GeV$^2$,
the increase is not present at this scale. 
In both cases, the dependence of the ratio on $x_2$
is quite moderate for $x_2>10^{-2}$.}
\label{modeldep1}
\end{figure}

\begin{figure}
\centerline{\psfig{file=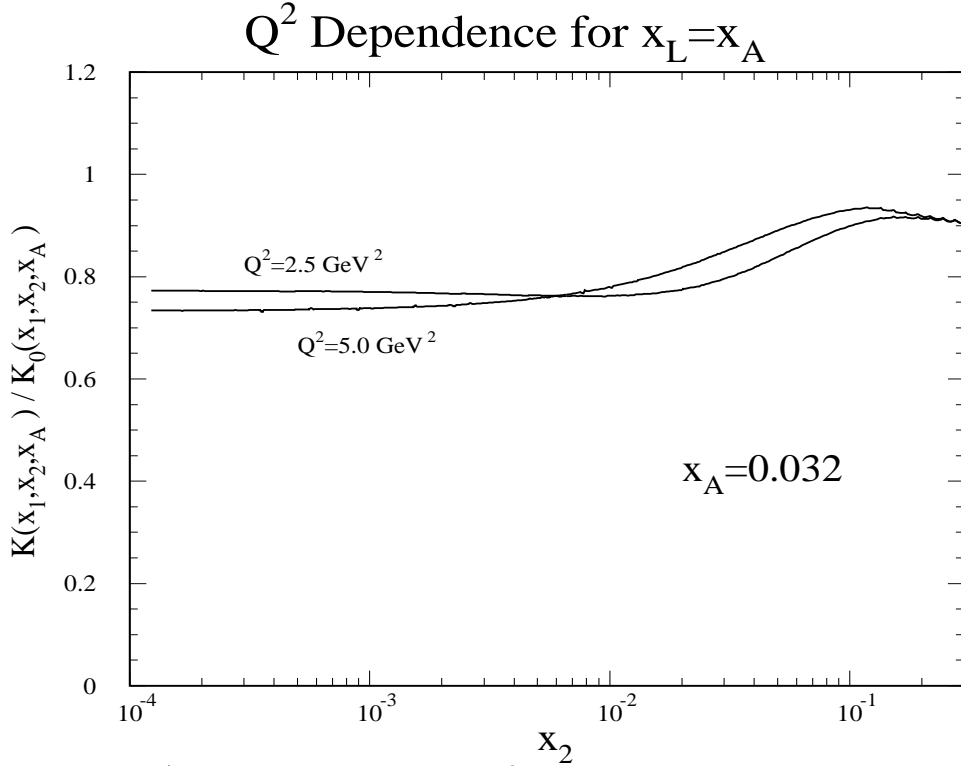,width=5.0in,height=4.0in}}
\caption{When $x_L\neq0$, the wild dependence of the ratio
on $x_2$ is suppressed as $x_2$ decreases.  This leads
to much more moderate behavior over the whole range of $x_2$.}
\label{modeldep2}
\end{figure}

To see this explicitly, we plot the ratio $K/K_0$ as a function
of $x_2$ in Figure \ref{modeldep1}.  Here, we can clearly see that the 
large variations are confined to $x_2\lsim0.01$.  Changing
the scale of the input distributions allows us to explore the 
behavior of our ratio as a function of $Q^2$.  In particular,
the turnover of the gluon momentum distribution at $Q^2=5.0$ GeV$^2$
occurs below our cut-off at $x=10^{-4}$.  Hence the 
increase observed at $Q^2=2.5$ GeV$^2$ is no longer present.
Instead, we see a decrease induced by the increasing value of the 
gluon momentum distribution.  The numerator is approximately
constant in the region of small $x_2$ due to the smearing of the 
convolution.
Since our ratio actually depends on the combination 
$x_2+x_L/2$ rather than $x_2$, the small-$x_2$ behavior is
stabilized by taking $x_L$ finite.  As displayed in 
Figure \ref{modeldep2}, our ratio saturates as $x_2$ 
is reduced when $x_L$ is finite.  In addition, its
behavior is far more moderate. 

Our numerical analysis has shown that in a very
general region the `naive' expectation, Eq.(\ref{naiveexp}), for our
matrix element is quite a good approximation.
As long as $x_A$ and $x_L/x_A$ are small enough, $x_2+x_L/2$
is not too small, and  $0.2\lsim x_1\lsim 0.7$, 
corrections run in the 10-15\% level
or less.  We can get even better approximations by
including an $x_A$-dependent constant of order $0.8$.
However, we must be very careful when using this approximation
generally.  It is easy to see from the above plots that the 
ratio varies quite quickly as one leaves the region of 
validity.

\section{Phenomenology}

In previous studies \cite{gq}, one has assumed a Gaussian spatial 
nuclear distribution $\exp[-(y^-/2R^-_A)^2]$, which leads to
a phenomenological form for the two-parton correlation in nuclei,
\begin{equation}
T^A_{qg}(x,x_T,x_L)=\widetilde{C}MR_A AQ(x) (1-e^{-x_L^2/x_A^2}).
\label{eq:tqg2}
\end{equation}
Compared with the recent HERMES experimental data \cite{hermes} on
nuclear modification of quark fragmentation function in DIS, one
has extracted \cite{ww02} 
the constant $\widetilde{C}\alpha_s^2\approx 0.00065$ GeV$^2$.
The strong coupling constant $\alpha_s$ should be evaluated at
a scale $Q^2\approx 2.8$ GeV$^2$. Such a value is also consistent with
what is extracted from the nuclear transverse momentum broadening \cite{gq},
$\widetilde{C}\alpha^2_{\rm s}=0.00086$ GeV$^2$.

Assuming now that $K_0(x_1,x_2,x_L)$ is a good approximation of 
$K(x_1,x_2,x_L)$, we can express
the two-parton correlation in nuclei of Eq.(\ref{TKsim}) as
\begin{eqnarray}
T^A_{qg}(x,x_T,x_L)=\frac{A^2}{32R_A^2}\langle R_A^2\Delta_\perp^2\rangle
& &\left[Q(x+x_L)G(x_T)+Q(x)G(x_T+x_L) \right. \nonumber\\
& & \left. -2Q(x+x_L/2)G(x_T+x_L)e^{-\xi^2} \right].
\end{eqnarray}
Here we assume a Gaussian distribution for 
the two-nucleon correlation function $\tilde r_2$. For moderate
values of $x>\!\!>x_L\lsim x_A<\!\!<1$, 
one can assume $Q(x+x_L)\approx Q(x)$, the above expression can
be approximated as
\begin{eqnarray}
T^A_{qg}(x,x_T,x_L)&=&\frac{A^2}{32R_A^2}\langle R_A^2\Delta_\perp^2\rangle
Q(x)\left[G(x_T)+G(x_T+x_L)\right] \nonumber\\
& \times&\left[1-\frac{2G(x_T+x_L/2)}{G(x_T)+G(x_T+x_L)}
 e^{-x_L^2/4x_A^2}\right].
\label{factk}
\end{eqnarray}

The above derived factorization form is very close to
the phenomenological one in Eq.(\ref{eq:tqg2}), especially for
$x_L\ll x_T$. For large $x_L\gg x_T$, when $G(x_T)\gg G(x_T+x_L)$,
the coefficient in front of the exponential factor will have additional
suppression as compared to the phenomenological model. However, for not 
so large $Q^2$ and $x_L\lsim x_A$, 
$G(x_T) \sim  G(x_T+x_L)$. In this kinematic region, one can 
then relate the parameters in the phenomenological form to our 
result within the convolution model
\begin{equation}
\widetilde{C}MR_A \simeq \frac{A}{4R_A^2}
\langle R_A^2\Delta_\perp^2\rangle G(x_T)\;\; .
\end{equation}
Here, we have reduced the nuclear radius by 1/2 in Eq.~(\ref{factk})
in order to match the phenomenological form in Eq.~(\ref{eq:tqg2}).
With the value of $\widetilde{C}\alpha_s^2\approx 0.00065$ from
HERMES experiment and $G(x_T)\approx 3 $ at 
$x_T=x_B \langle k_T^2\rangle/Q^2\simeq 0.01$ 
($x_B=0.124$, $Q^2=2.8$ GeV$^2$, $\langle k_T^2\rangle\simeq 0.25$ GeV$^2$),
we have
\begin{equation}
\langle R_A^2\Delta_\perp^2\rangle \simeq 1.65,
\end{equation}
which is on the order of 1 and independent of the $A$ as expected.
One can consider this as a qualitative agreement between our calculated
two-parton correlation in nuclei and the experimental measurements.

\section{Conclusion and Discussion}

In this paper, we have studied the generalized twist-four nuclear 
parton matrix elements $K(x_1,x_2,x_L)$, 
both direct and momentum-crossed ones,
in the framework of a Fock hadronic state expansion of the nuclear states.
These twist-four nuclear parton matrix elements determine the effect of
multiple parton scattering in hard processes involving nuclei,
{\it e.g.}, the nuclear modification of the fragmentation functions.
Assuming that the contributions of higher Fock states induced by nucleonic
interaction are small, we have shown that the leading contribution
to the twist-four nuclear parton matrix elements can be expressed as
a convolution of twist-two nucleonic off-forward parton
distributions and the two-nucleon correlation function inside a 
nucleus. In the limit of extremely large nuclei, $A\rightarrow \infty$,
or $x_A=1/MR_A\rightarrow 0$,
we have also shown that the twist-four nuclear parton matrix
elements can factorized into the product of twist-two nucleonic
parton distributions. However, we demonstrated that the
matrix elements $K$ are not analytic in $x_A$ at $x_A=0$
(Corrections around $x_A=0$ cannot be expanded as powers of $x_A$).

To verify the factorization approximation, we have evaluated
the twist-four nuclear matrix elements numerically as the
convolution of twist-two nucleonic parton distributions and
two-nucleon correlation functions inside a nucleus, 
assumed to have a simple Gaussian form. For $x_L\lsim x_A$
and moderate $x_1$, we found that the factorization is a good
approximation within $\sim$20\% for large nuclei. However, the deviations
become very significant for small $x_{1}$ or $x_2+x_L/2$, large $x_A$,
or $x_L\gsim x_A$.
The corrections at small $x_2+x_L/2$ are particularly large when the gluon
momentum distribution $G(x_2+x_L/2)$ 
is large. Furthermore, for $x_{1,2}\ll x_L/2$,
one can no longer express the nuclear matrix elements as the convolution
of twist-two nucleonic parton distributions. In this region, 
off-forward nucleonic 
parton distributions, which are not known experimentally,
are needed for the numerical evaluation.
Therefore, one could conceivably use the measured 
nuclear effects in this kinematic region to constrain
the nucleonic OFPD's.

Another important nuclear effect on the parton matrix elements
that we have not considered so far is similar to the nuclear 
shadowing of the parton distributions or depletion of the effective
parton distribution per nucleon. 
This effect is known experimentally to be large for small $x_B$,
at least for the quark distributions \cite{nmc}.
One can understand nuclear shadowing as the consequence
of coherent initial scattering processes involving multiple 
nucleons \cite{strikman,huang} when one calculates the nuclear
parton distributions in terms of nucleonic parton distributions. 
In the framework
of our lowest Fock state expansion, these multiple scattering
processes will involve direct multi-nucleonic correlations.
Therefore, such multiple scattering
effects in principle are higher-twist contributions and should be 
suppressed at very large $Q^2$. 
In our calculation of the twist-four nuclear parton matrix
elements, such higher-twist contributions from multiple
scattering processes involving more than two nucleons 
should also contribute, but they are suppressed at large
$Q^2$.

\acknowledgements
The work is supported in part by
the Director, Office of Science, Office of High Energy and Nuclear
Physics, and by the Office of Basic Energy Sciences, 
Division of Nuclear Sciences of  
the U.S.~Department of Energy under Grant No. 
DE-AC03-76SF-00098 and in part by NSFC under project No. 19928511

\end{document}